\documentclass[english,aps,prx,showpacs,citeautoscript,twocolumn]{revtex4}
\pdfoutput=1
\usepackage{amsmath}
\usepackage{amssymb}
\usepackage{graphicx}
\usepackage{array}
\usepackage{float}
\usepackage{braket}

\date{\today}

\begin{document}
\title{Multiferroic crossover in perovskite oxides}

\author{L.\ Weston,$^1$ X.\ Y.\ Cui,$^{2, 3}$ S.\ P.\ Ringer,$^{2, 3}$ and C.\ Stampfl$^1$}
\affiliation{$^1$School of Physics, The University of Sydney, Sydney, New South Wales,
2006, Australia}
\affiliation{$^2$Australian Centre for Microscopy and Microanalysis, The University of Sydney, New South Wales, 2006, Australia}
\affiliation{$^3$School of Aerospace, Mechanical and Mechatronic Engineering, The University of Sydney, New South Wales, 2006, Australia}

\begin{abstract}
The coexistence of ferroelectricity and magnetism in ABO$_3$ perovskite oxides is rare, a phenomenon that has become known as the ferroelectric ``$d^0$ rule". Recently, the perovskite BiCoO$_3$ has been shown experimentally to be isostructural with PbTiO$_3$, while simultaneously the $d^6$ Co$^{3+}$ ion has a high spin ground state with $C$-type antiferromagnetic ordering. It has been suggested that the hybridization of Bi 6$s$ states with the O $2p$ valence band stabilizes the polar phase, however we have recently demonstrated that Co$^{3+}$ ions in the perovskite structure can facilitate a ferroelectric distortion via the Co 3$d$ -- O $2p$ covalent interaction (Weston \textit{et al.}, Phys. Rev. Lett. \textbf{114}, 247601 (2015)).
In this report, using  accurate hybrid density functional calculations, we investigate the atomic, electronic and magnetic structure of BiCoO$_3$ to elucidate the origin of the multiferroic state. To begin with, we perform a more general first-principles investigation of the role of $d$ electrons in affecting the tendency for perovskite materials to exhibit a ferroelectric distortion; this is achieved via a qualitative trend study in artificial cubic and tetragonal LaBO$_3$ perovskites. We choose La as the A-cation so as to remove the effects of Bi $6s$ hybridization. The lattice instability is identified by the softening of phonon modes in the cubic phase, as well as by the energy lowering associated with a ferroelectric distortion. For the LaBO$_3$ series, where B is a $d^0 - d^8$ cation from the $3d$-block, the trend study reveals that increasing the $d$ orbital occupation initially removes the tendency for a polar distortion, as expected. However, for high spin $d^5-d^7$ and $d^8$ cations a strong ferroelectric instability is recovered. This effect is explained in terms of increased pseudo Jahn-Teller (PJT) $p - d$ vibronic coupling. The PJT effect is described by the competition between a stabilizing force ($K_0$) that favours the cubic phase, and a vibronic term that drives the ferroelectric state ($K_v$). The recovery of the lattice instability for high spin $d^5-d^7$ and $d^8$ cations is due to (i) a reduction in $K_0$ due to a significant volume increase arising from population of the $\sigma$-bonded axial $d$ $e_g$ orbitals, and (ii) an increase in the $K_v$ contribution arising from increased $p-d$ hybridization; our calculations suggest that the former mechanism is dominant. Surprisingly, we are able to show that, in some cases unpaired electron spins \emph{actually drive ferroelectricity}, rather than inhibit it, which represents a shift in the understanding of how ferroelectricity and magnetism interact in perovskite oxides. It follows, that for the case of BiCoO$_3$, the Co$^{3+}$ ion plays a major role in the ferroelectric lattice instability. Importantly, the ferroelectric polarization is greatly enhanced when the Co$^{3+}$ ion is in the high spin state, when compared to the nonmagnetic, low spin state, and a large coupling of the electric and magnetic polarization is present. 
Generally, for $d^5-d^7$ B-cations in ABO$_3$ perovskites, an inherent and remarkably strong magnetoelectric coupling exists via the multiferroic crossover effect, whereby switching the spin state strongly affects the ferroelectric polarization, and potentially, manipulation of the polarization with an externally applied electric field could induce a spin state transition. This novel effect is demonstrated for BiCoO$_3$, for which the ground spin state is switched by reducing the internal ferroelectric polarization. These results provide a deeper insight into perovskite ferroelectrics and multiferroics.
\end{abstract}


\maketitle
\def\figurename{FIG.}\def\tablename{TABLE}

\section{Introduction}

The apparent mutual exclusion of ferroelectricity and magnetism in ABO$_3$ perovskite oxides has both fascinated and puzzled researchers in the field of multiferroics for decades  \cite{HillJPC2000, KhomskiiJMMM2006}. It is generally known that the presence of $d$-electrons at the B-site cation, which is a requirement for magnetism, reduces or removes the tendency for noncentrosymmetric distortions of the BO$_6$ octahedra, a phenomenon which has been dubbed the ``$d^0$ rule". While there has been much progress in this field, the observed mutual exclusion between ferroelectricity and magnetism remains to be fully understood \cite{KhomskiiPHY2009}.

Perovskite oxides undergo various symmetry lowering distortions which decrease the total energy of the system. In the case of a ferroelectric material, the noncentrosymmetric distortions of BO$_6$ octahedra leads to breaking of inversion symmetry and a net macroscopic electric polarization  \cite{CohenNAT1992}. It is more common, however, for ABO$_3$ perovskites to undergo centrosymmetric distortions involving the tilting and rotation of BO$_6$ octahedra \cite{BenedekJPC2013}. It has been suggested that competition with such centrosymmetric distortions precipitates the ferroelectric $d^0$ rule \cite{HillJPC2000}. 

Ferroelectricity in perovskite oxides can have different origins, e.g., from canted spins in frustrated magnets \cite{CheongNAT2007}, or from size and electrostatic effects in the hexagonal phase \cite{VanAkenNAT2004}; however, in this work, by using the term ferroelectricity we are implicitly referring to ``proper'' ferroelectricity. Here, $p-d$ covalent bonding between the B-cation $d$ states and surrounding O $2p$ states can stabilize the ferroelectric phase, which is characterized by a shift of the B-cation sublattice with respect to that of the surrounding O ions \cite{CohenNAT1992}. Examples of proper ferroelectrics include the classic cases of BaTiO$_3$ and PbTiO$_3$, where the Ti$^{4+}$ ion does indeed have a $d^0$ electron configuration, and therefore both BaTiO$_3$ and PbTiO$_3$ are non-magnetic ferroelectric perovskites, consistent with the $d^0$ rule \cite{HillJPC2000}. It has recently, however, been pointed out by Bersuker \cite{BersukerPRL2012}, that certain $3d^n$ ($n$ $>$ 0) magnetic cations should be unstable against a pseudo Jahn-Teller (PJT) driven ferroelectric distortion. This proposition suggests that there exists no fundamental explanation for the ferroelectric $d^0$ rule within the PJT theory for ferroelectricity, although an examination of experimental results suggests, at least empirically, that magnetic B-cations (almost) never exhibit proper ferroelectricity. Nevertheless, currently there is a strong research effort focused on the design of        magnetoelectric multiferroics, as the coupling of ferroelectricity and magnetism within a single phase material promises novel and exciting opportunities for future spintronic devices  \cite{NanJAP2008, RoyACMP2012, GajekNAT2007, YangJAP2007, BibesNAT2008, ScottNAT2007}.

The perovskite oxide BiCoO$_3$ has been shown to exhibit a tetragonal ferroelectric phase as the ground state (i.e., isostructural with PbTiO$_3$) \cite{BelikCM2006}. At the same time, the $d^6$ Co$^{3+}$ ion is in the high spin state, with $C$-type antiferromagnetic ordering. This system is therefore a rare example of a $d^n$ perovskite exhibiting proper ferroelectricity, and represents an opportunity to to investigate the interaction of ferroelectricity and magnetism in perovskite oxides. It has been proposed, based on first-principles calculations \cite{JiaPRB2012}, that the ferroelectric phase of BiCoO$_3$ is driven by hybridization of the Bi 6$s$ states and the O $2p$ valence band; however, we recently demonstrated that Co$^{3+}$ ions doped into PbTiO$_3$ exhibit novel ferroelectric and multiferroic effects \cite{WestonPRL2015}, and we therefore propose that the Co$^{3+}$ ion is responsible for the strong ferroelectric distortions exhibited by BiCoO$_3$ \cite{BelikCM2006}, as well as for the multiferroic effects predicted theoretically \cite{RavindranAM2008}.

At this point, it is clear that more work is needed, both experimental and theoretical, so as to understand novel multiferroic materials such as BiCoO$_3$, as well as to further elucidate the complex interactions of ferroelectricity and magnetism on a microscopic scale. In this report, the interaction of ferroelectricity and magnetism in ABO$_3$ perovskites is investigated in detail. It is demonstrated that the presence of $d$-electrons at the B-cation does not necessarily rule out a strong ferroelectric distortion; in fact, in the case of BiCoO$_3$, the presence of unpaired $d$-electrons actually \emph{drives} ferroelectricity, rather than inhibits it. The manuscript is presented as follows: the methodology is described in section~\ref{sec:methods} and our results and discussion are presented in section~\ref{sec:results}. In section~\ref{sec:PJTE}, we provide a brief introduction into perovskite ferroelectricity and multiferroicity. Section~\ref{sec:LaBO3} investigates the interaction of ferroelectricity and magnetism via a qualitative trend study for artificial LaBO$_3$ perovskites, and in section~\ref{sec:MFCE} we describe quantitatively the origins of ferroelectricity and multiferroicity for the realistic perovskite BiCoO$_3$. Our key findings are summarized in section~\ref{sec:summary}.

\section{Methodology} 
\label{sec:methods}
Calculations are performed within the density functional theory (DFT), using the screened hybrid functional of Heyd, Scuseria and Ernzerhof (HSE) \cite{HeydJCP2003} for the exchange-correlation potential. We use predominantly the standard implementation HSE06, where the screening length and mixing parameter are fixed to 10 \AA \ and 0.25 respectively. While the inclusion of 25\% Hartree-Fock in the calculation of the exchange potential gives a highly accurate account of the atomic and electronic structure \cite{MarsmanJPC2008}, for the calculations in Fig.~\ref{fig:SWITCH}, the mixing parameter is reduced to 0.125 so as to give a more accurate description of the spin splitting energy \cite{ReiherTCA2001, HarveyBOOK2004}. The valence electrons are separated from the core by use of the projector augmented wave (PAW) \cite{BlochlPRB1994} pseudopotentials as implemented in the VASP package \cite{KressePRB1996}. The energy cutoff for the plane wave basis set is 500 eV. For the 5-atom LaBO$_3$ primitive cells a 6$\times$6$\times$6 \textbf{\textit{k}}-point grid is used for integrations over the Brillouin zone. For the 20-atom BiCoO$_3$ cells a 4$\times$4$\times$3 grid is used. Full optimization of the lattice vectors is allowed, and the internal coordinates are relaxed until the forces are less than 0.001 eV/\AA. The spin state of each perovskite is controlled using fixed-spin moment calculations. The phonon frequencies are calculated using the method of finite differences. As has been previously reported for phonon calculations with DFT-VASP \cite{EvarestovPRB2011}, we find that large values for the displacements are required to obtain reasonable results, and we use displacements of 0.04 in lattice units; displacements this large are expected to be outside of the linear regime, however we also present the results of total energy calculations and lattice relaxations to confirm the findings of the phonon study.

\section{Results and Discussion}
\label{sec:results}

\subsection{Perovskite Ferroelectricity and the Pseudo Jahn-Teller Effect}
\label{sec:PJTE}

\begin{figure}
\includegraphics[width=8.5cm]{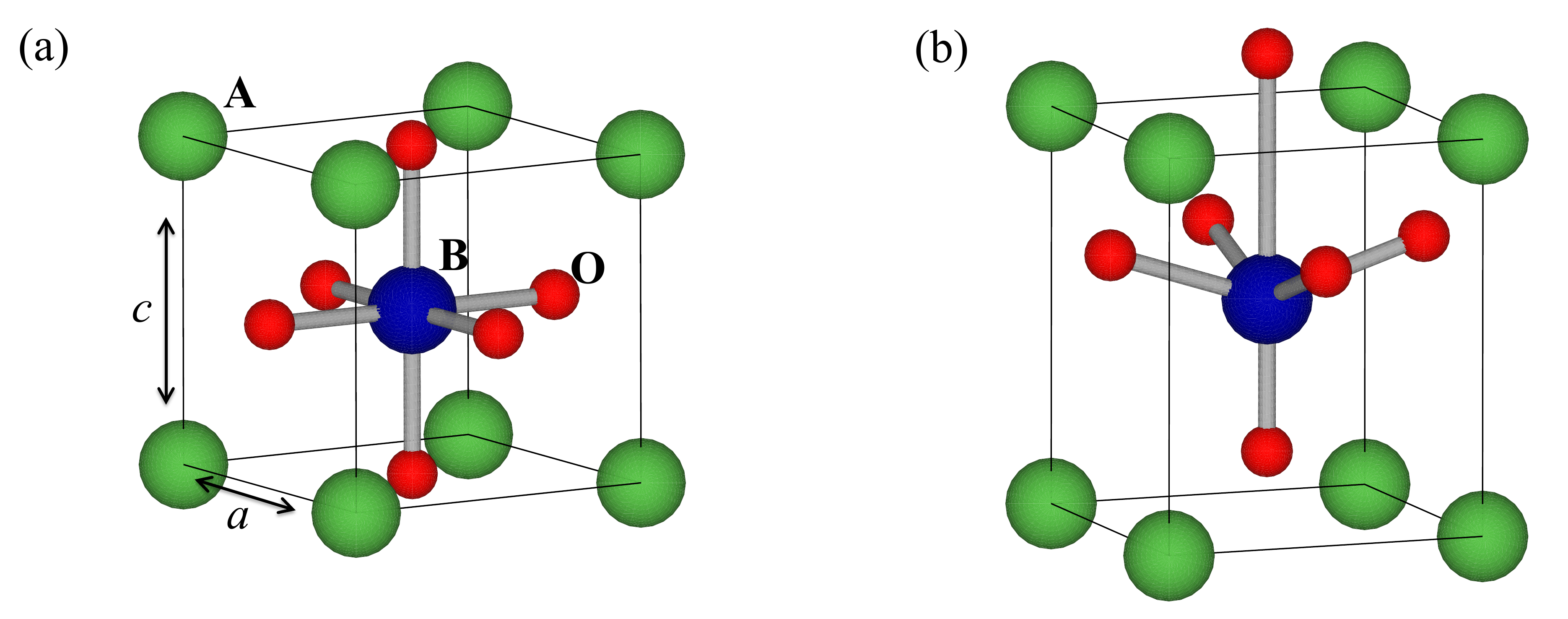}
\caption{(Color online) (a) The unit cell of an ABO$_3$ perovskite oxide with a cubic crystal structure. The B-cation is at the centre of the unit cell and is 6-fold coordinated to oxygen. (b) The same system with a tetragonal ferroelectric distortion. The distortion is characterized by an elongation of the $c$-axis compared to the in-plane lattice vectors, as well as a shift of the B-cation with respect to the surrounding O-ion octahedron; the internal distortions break inversion symmetry, leading to a net macroscopic electrical polarization. The A-cation of the perovskite structure is represented by a green sphere, the B-cation is blue, and O ions are red.}
\label{fig:ABO3}
\end{figure}

\begin{figure}
\includegraphics[width=8.7cm]{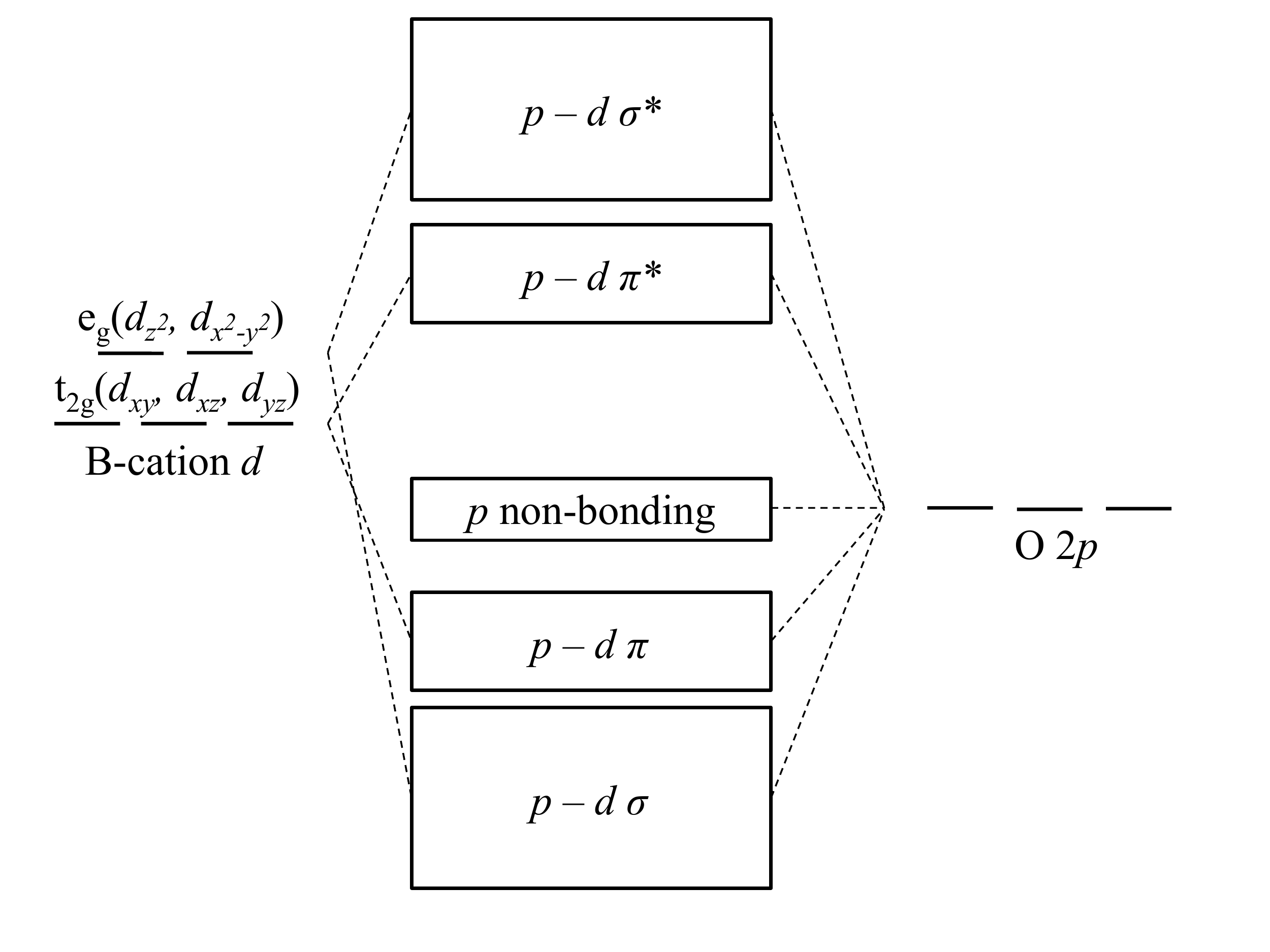}
\caption{(Color online) A molecular orbital diagram for an ABO$_3$ perovskite. The bonding states ($\pi$ and $\sigma$) have predominant O $2p$ character, the antibonding states ($\pi^*$ and $\sigma^*$), have predominant B-cation $d$ character. For simplicity, only the valence states are included, and the A-cation orbitals have been neglected.}
\label{fig:MO}
\end{figure}

To begin with, we describe the origins of perovskite ferroelectricity in terms of the PJT theory. The cubic crystal structure of an ABO$_3$ perovskite oxide is shown in Fig.~\ref{fig:ABO3}(a). The A-cation is 12-fold coordinated to oxygen, and generally, the interaction with the surrounding O ions is ionic, in that the A-cation valence electrons are transferred to the O ion $2p$ band without significant hybridization. On the other hand, the B-cation (shown here at the center of the unit cell) is 6-fold coordinated to oxygen, and the B-cation $d$ states will generally show a significant covalent interaction with the O $2p$ states. A basic molecular orbital diagram for this covalent interaction is shown in Fig.~\ref{fig:MO}. The O $2p$ and B-cation $d$ states will form bonding and antibonding states, with the former having predominant O $2p$ character, and the latter having predominant B-cation $d$ character. Due to the octahedral crystal field, the free ion $d$ band is split into a $t_{2g}$ triplet and an $e_g$ doublet, for which the $p-d$ interaction is $\pi$-bonding and $\sigma$-bonding, respectively.

Perovskite oxides can deviate from the ideal cubic geometry presented in Fig.~\ref{fig:ABO3}(a). For the case of a proper ferroelectric distortion, the B-cation at the center of the cubic unit cell will shift with respect to the surrounding O-ion octahedron, leading to a net electrical polarization, as shown in Fig.~\ref{fig:ABO3}(b). The driving force for the lattice instability is the   PJT effect \cite{BersukerBOOK2006, BersukerCR2013, PolingerPB2015}, and this phenomenon can lead to coupled magnetic and dielectric polarizations \cite{BersukerPRL2012, GarciaPRL2011, GarciaPRB2011}; here, we present the key points of the theory. For a perovskite oxide originally in the cubic (paraelectric) phase, for a symmetry-lowering ferroelectric distortion, the adiabatic potential energy surface $E(q)$ along the ionic displacement $q$ has a curvature $K$,

\begin{eqnarray}
K = \left(\frac{\delta^2 E}{\delta q^2}\right)_0 \quad. \label{curvature}
\end{eqnarray}

\noindent The lattice exhibits an instability if the curvature at the high symmetry point is negative, $K < 0$. As $E = \Braket{\Psi_0 | H | \Psi_0}$, where $H$ is the Hamiltonian and $\Psi_0$ is the corresponding wavefunction in the ground state, Eq.~\ref{curvature} can be expanded,

\begin{eqnarray}
K = \Braket{\Psi_0 |\left(\frac{\delta^2 H}{\delta q^2}\right)_0 | \Psi_0} + 2\Braket{\Psi_0 |\left(\frac{\delta H}{\delta q}\right)_0 | \Psi_0'}, \label{PJTE}
\end{eqnarray}
\noindent where, $\Psi_0' = \left(\frac{\delta \Psi_0}{\delta q}\right)_0$. From Eq.~\ref{PJTE}, it is clear that $K$ is the sum of two competing terms,

\begin{eqnarray}
K = K_0 + K_v \quad, \label{PJTEsum}
\end{eqnarray}

where,

\begin{eqnarray}
K_0 = \Braket{\Psi_0 |\left(\frac{\delta^2 H}{\delta q^2}\right)_0 | \Psi_0} \quad.\label{PJTErestoring}
\end{eqnarray}

\noindent Within second order pertubation theory, $K_v$ is expressed as follows \cite{BersukerBOOK2006, BersukerPRL2012},

\begin{eqnarray}
K_v = \sum_n \frac{\left|\Braket{\Psi_0 | \left(\frac{\delta H}{\delta q}\right)_0 | \Psi_n}\right|^2}{(E_0 - E_n)} \quad.  \label{PJTEvibronic} 
\end{eqnarray}

\noindent Here, $\Psi_0$ is the ground state wavefunction of the unperturbed Hamiltonian with energy $E_0$, and $\Psi_n$ are excited states with energy $E_n$. The PJT effect provides the driving force for ferroelectric distortions in perovskite oxides \cite{BersukerPRL2012}. Whether or not a polar ferroelectric phase is favoured (i.e., a negative curvature $K$) depends on the relative strengths of the contributions $K_{0}$ and $K_{v}$ in Eq.~\ref{PJTEsum}. The term $K_{0}$ described by Eq.~\ref{PJTErestoring} is always positive -- as the wavefunction $\Psi_0$ represents the ground electronic state for the unperturbed Hamiltonian, moving the coordinates along $q$ will always increase the total energy. Therefore $K_{0}$ provides a restoring force that favours the undistorted phase. The term $K_{v}$ is the so-called ``vibronic term", which lowers the total energy as it corresponds to the response of the electronic wavefunction to the distortion, which is the mixing of $\Psi_0$ with appropriate excited states, $\Psi_n$. The matrix elements $\braket{\Psi_0 |  (\delta H/\delta q)_0 | \Psi_n }$ in Eq.~\ref{PJTEvibronic} are only non-zero if $\Psi_0$ and $\Psi_n$ have the same spin, and due to the denominator ($E_0-E_n$) it is clear that a strong polar distortion requires that low-lying excited states exist, with the same spin-multiplicity as the ground state. For PJT driven ferroelectricity, one can envision three possible scenarios: (i) when $K_{0} \gg K_{v}$, the system will be stable against a ferroelectric distortion, (ii) when $K_{0} \sim K_{v}$, the two phases have a similar energy, and (iii) when $K_{v} \ll K_{v}$, the system will exhibit a strong lattice instability that favours the polar phase.

For the case of perovskite oxides, the vibronic term of Eq.~\ref{PJTEvibronic} describes how a ferroelectric distortion is driven by increased overlap of O $2p$ states with the $d$ states of the B-cation \cite{BersukerPRL2012, BersukerJP2012}. From Eq.~\ref{PJTEvibronic}, it is clear that the strength of the vibronic coupling between a $p$ and $d$ orbital depends on, firstly, how strongly the overlap of the orbitals is increased by the distortion, $\braket{p |  (\delta H/\delta q) | d}$, and secondly, the energy gap between the two states ($E_p - E_d$). In principle, the O $2p$ states can form covalent bonds with the non-axial $t_{2g}$ orbitals, or the axial $e_g$ orbitals. The $t_{2g}$ states lie at a lower energy than the $e_g$ states, and so according to the denominator in Eq.~\ref{PJTEvibronic}, should lead to a greater hybridization with O $2p$ states, suggesting the $\pi$-bonding $p-t_{2g}$ interaction is dominant in driving ferroelectricity; however, both the $d$ $e_g$ states and the O $2p$ states are axial in nature, and therefore have a greater potential for overlap (and hence covalency), and therefore the $\sigma$-bonding $p-e_{g}$ interaction is also important.

\subsection{The LaBO$_3$ Series}

\label{sec:LaBO3}

The key goal of this study is to determine the microscopic origins of the ferroelectric and multiferroic behaviour of BiCoO$_3$; however, based on our previous work \cite{WestonPRL2015}, we propose that the Co$^{3+}$ ion drives the lattice instability, rather than (or in conjunction with) the Bi 6$s$ hybridization. To isolate the contribution of the B-cation to the ferroelectric lattice instability in ABO$_3$ perovskites, we perform a trend study of the LaBO$_3$ series (ruling out contributions from Bi). A comparison between the contributions of Bi and La to the ferroelectric instability of perovskites has been previously reported \cite{HillJPC2000}; additionally, we will further compare the role of the A-cation in stabilizing the polar phase for LaBO$_3$ and BiBO$_3$ perovskites in section~\ref{sec:MFCE}. For the remainder of this section, we focus on the role of the B-cation in the lattice instability of LaBO$_3$ perovskites.

As has been discussed above, perovskites will typically undergo a competing centrosymmetric distortion, rather than a polar one; this is true across the entire LaBO$_3$ series. In Table~\ref{tab:structure}, we present the experimentally reported ground state phases for each LaBO$_3$ perovskite. As can be seen, the LaBO$_3$ perovskites typically prefer to undergo orthorhombic and rhombohedral distortions involving the tilting and rotation of BO$_6$ octahedra, rather than exhibiting polar ferroelectric distortions. This fact, and the general lack of magnetic ferroelectric perovskites \cite{HillJPC2000}, necessitates the study of artificial systems so as to perform trend studies that can allow insight into the ferroelectric $d^0$ rule, and perovskite multiferroics. 
Historically, one of the key ways in which multiferroicity has been studied, is to take a magnetic perovskite like the LaBO$_3$ systems presented in Table~\ref{tab:structure}, and to calculate the ferroelectric lattice instability in the cubic phase, e.g., by analysis of phonon softening \cite{HillJPC2000, HongPRB2012}. For example, analysis of the spin -- phonon coupling in the cubic phase of LaMnO$_3$ and LaFeO$_3$ has been used to reveal multiferroic effects \cite{HongPRB2012}. With this same approach, below, we perform a qualitative trend study to examine the ferroelectric lattice instability in the LaBO$_3$ perovskites as a function of the $d$ orbital occupation and configuration, so as to provide insight into perovskite ferroelectricity and possible multiferroicity; for simplicity, to begin with, we neglect the long range inter-ionic magnetic ordering for the LaBO$_3$ systems as we try to elucidate qualitative trends. However, in the next section, we present quantitative calculations focusing on magnetoelectric coupling for the realistic perovskite system BiCoO$_3$.

\begin{table}
\setlength{\tabcolsep}{10pt}
\setlength{\extrarowheight}{4.5pt}
\begin{tabular}{lcc} \hline\hline
 Perovskite & Crystal structure & Reference \\ \hline
LaScO$_3$ & Orthorhombic  & [\onlinecite{LiferovichJSSC2004}]  \\ 
LaTiO$_3$ & Orthorhombic  & [\onlinecite{CwikPRB2003}]  \\  
LaVO$_3$  & Monoclinic  & [\onlinecite{RenPRB2003}] \\ 
LaCrO$_3$ & Orthorhombic &  [\onlinecite{CarterJMS1996}] \\  
LaMnO$_3$  & Orthorhombic & [\onlinecite{RodriguezPRB1998}]  \\ 
LaFeO$_3$ & Orthorhombic  & [\onlinecite{MarezioMRB1971}]  \\  
LaCoO$_3$  & Rhombohedral & [\onlinecite{AsaiJPSJ1998}]  \\ 
LaNiO$_3$ & Rhombohedral &  [\onlinecite{GarciaPRB1992}] \\  
LaCuO$_3$  & Rhombohedral & [\onlinecite{WebbPLA1989}]  \\   \hline\hline
\end{tabular} 
\caption{Low temperature phases of the LaBO$_3$ perovskites in the ground state.}
\label{tab:structure}
\end{table}

\begin{figure}
\includegraphics[width=8cm]{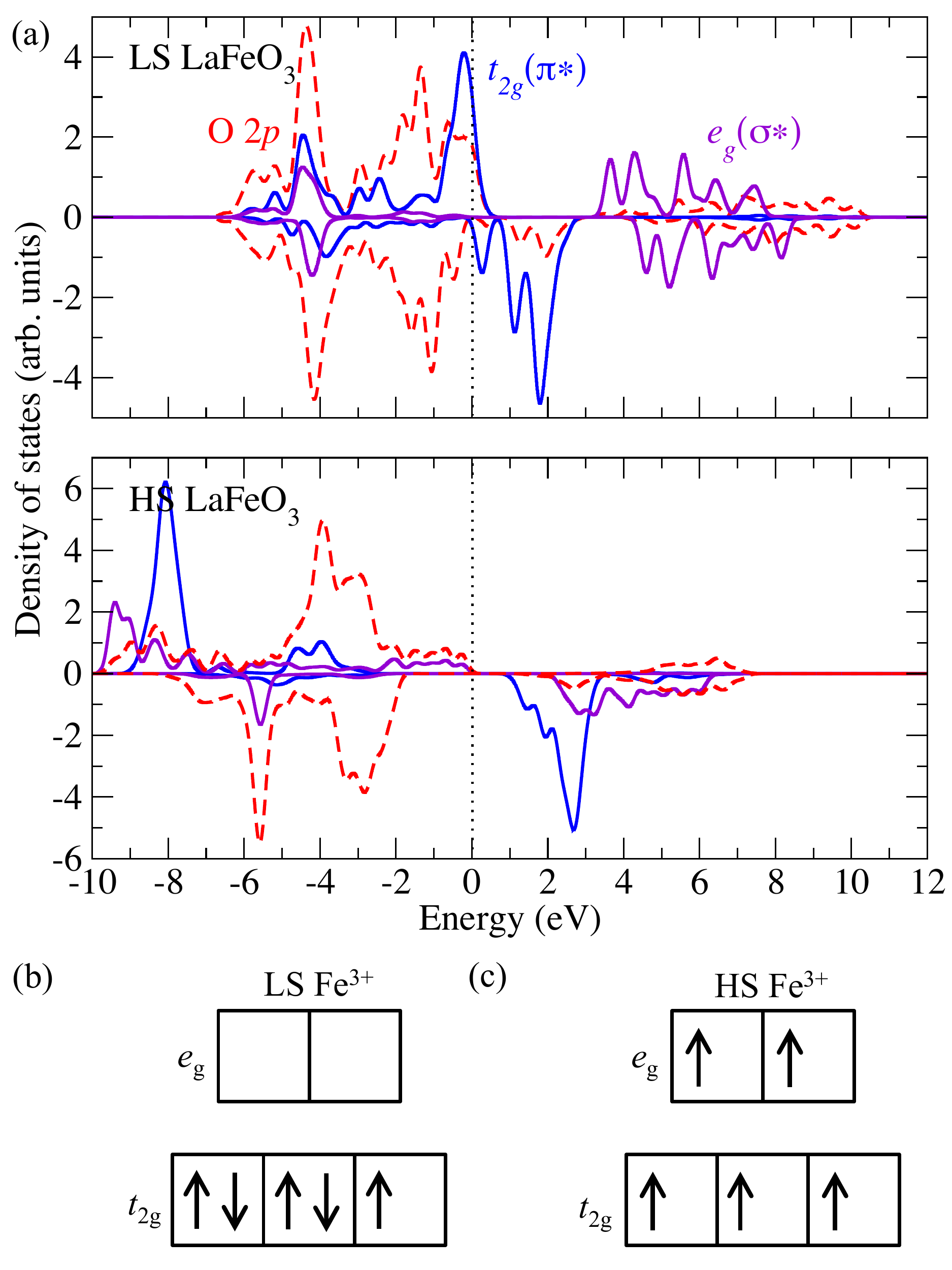}
\caption{(Color online) In (a), the projected partial density of states is plotted for low spin (LS) (top) and high spin (HS) (bottom) LaFeO$_3$. The dotted red line represents O $2p$ states, the solid blue and purple lines represent respectively the $t_{2g}$ ($d_{xy}$, $d_{xz}$, $d_{yz}$) and $e_g$ ($d_{z^2}$, $d_{x^2-y^2}$) orbitals of the Fe $3d$ band. In (b) and (c), the $d$ orbital configurations are shown for LS and HS Fe$^{3+}$.}
\label{fig:pdos}
\end{figure}

In this section we study the LaBO$_3$ series, where B represents all of the possible B-cations (in the 3$+$ charge state \cite{noteCHARGE}) across the $3d$-series (i.e., B $=$ Sc$^{3+}$($d^0$), ..., Cu$^{3+}$($d^8$)). The electron configuration for each B-cation under study is presented in Table~\ref{tab:SLC2}. For $d^4-d^7$ B-cations, the octahedral crystal field splitting leads to two possible orbital configurations, known as the low spin (LS) and high spin (HS) states, which maximize and minimize spin pairing, respectively (in some cases an intermediate spin (IS) state is also possible). As representative examples of the electronic structure for the LaBO$_3$ systems, the density of states (DOS) is plotted for LS and HS LaFeO$_3$ in Fig.~\ref{fig:pdos}, as well as the expected orbital filling for the Fe $3d$ band. As can be seen, the DOS confirms the orbital fillings of Table~\ref{tab:SLC2}. For LS LaFeO$_3$, the $d^5$ configuration is such that only the $t_{2g}$ band is occupied, whereas the $e_g$ band is empty; on the other hand, in the HS state each $t_{2g}$ and $e_g$ orbital is singly occupied, and there exists a strong exchange-splitting, as well as increased $p-d$ hybridization.

As previously discussed, in the simple cubic perovskite structure, the B-cation is 6-fold coordinated to oxygen, in a configuration that is symmetrical with respect to inversion. We study the tendency for the LaBO$_3$ perovskite to exhibit a ferroelectric distortion, by calculating the  zone-center phonon modes for this high symmetry cubic phase --  a positive frequency indicates stability against a spontaneous distortion, an imaginary frequency, or soft mode, indicates that the system is unstable \cite{KingPRB1994}. Upon calculating the eigenvectors of the dynamical matrix, we identify the zone-center phonon mode that has been previously associated with ferroelectricity -- it is the softening of this mode that indicates a ferroelectric instability \cite{HillJPC2000}. An example of the eigenvector  for this mode, calculated for the HS state of LaFeO$_3$ is presented in Table~\ref{tab:mode}; our calculation closely reproduces the eigenvectors previously reported for LaBO$_3$ perovskites \cite{HillJPC2000}. In Table~\ref{tab:SLC2}, the frequencies of the ferroelectric mode ($\omega_{FE}$) are presented for each LaBO$_3$ system. As a further test, the lattice vectors and internal coordinates are presented for the LaBO$_3$ systems under study. The relaxations are started from a nearly cubic structure, however a small ferroelectric distortion is present. The systems that favour a ferroelectric distortion relax so as to increase the internal polarization and the $c/a$ axis ratio. On the other hand, systems which do not favour a ferroelectric distortion relax back to a paraelectric (cubic) structure, without internal distortions and with $c/a = 1$. We point out that for an orbitally degenerate configuration of the $d$-band, the system is also unstable against a Jahn-Teller driven distortion \cite{BersukerBOOK2006}; however, this is outside of the scope of the current work, and we present results for the cubic structure of LaBO$_3$ systems that cannot be stabilized in the PJT-driven ferroelectric phase.

\begin{table*}
\setlength{\tabcolsep}{4.8pt}
\setlength{\extrarowheight}{4.5pt}
\begin{tabular}{lcccccccccc} \hline\hline
B-cation & Spin state & Elec. Config. &  $\omega_{FE}$ (cm$^{-1}$)& $a$ (\AA) & $c$ (\AA)  & La & B & O1  & O2 & $\Delta E_{FE}$ (meV) \\ \hline
  Sc$^{3+}$ &   & $d^0$ &  155.7$i$ & 3.893 & 4.783  & 0.000 & 0.423 & 0.834 & 0.294 & $-327$ \\    
   Ti$^{3+}$ &   &  $d^1$; $t_{2g}(\uparrow)^1$ & 46.6$i$ & 3.950 & 3.959  & 0.000 & 0.456 & 0.952 & 0.437 & $-9$ \\ 
   V$^{3+}$ &   & $d^2$; $t_{2g}(\uparrow)^2$ & 3.0$i$	 & 3.863 &  4.023  & 0.000 & 0.484 & 0.983 & 0.474 & $-6$\\ 
   Cr$^{3+}$ &   & $d^3$; $t_{2g}(\uparrow)^3$ & 53.9 & 3.785 & 3.785 & 0.000 & 0.500 & 0.000 & 0.500 & - \\ 
 Mn$^{3+}$ & LS & $d^4$; $t_{2g}(\uparrow)^3t_{2g}(\downarrow)^1$ & 87.9 & 3.809 & 3.809 & 0.000 & 0.500 & 0.000 & 0.500 & -\\ 
  & HS & $d^4$; $t_{2g}(\uparrow)^3e_{g}(\uparrow)^1$ & 50.7 & 3.880 & 3.940  & 0.000 & 0.489 & 0.988 & 0.485 & $-9$ \\ 
  Fe$^{3+}$ & LS & $d^5$; $t_{2g}(\uparrow)^3t_{2g}(\downarrow)^2$ & 142.6 & 3.786 & 3.786 & 0.000 & 0.500 & 0.000 & 0.500 & - \\ 
  & IS & $d^5$; $t_{2g}(\uparrow)^3t_{2g}(\downarrow)^1e_{g}(\uparrow)^1$ & 80.9 & 3.839 & 3.839  & 0.000 & 0.500 & 0.000 & 0.500 & - \\
   & HS & $d^5$; $t_{2g}(\uparrow)^3e_{g}(\uparrow)^2$ & 87.8$i$ & 3.800 & 4.620  & 0.000 & 0.440 & 0.842 & 0.319 & $-76$ \\ 
  Co$^{3+}$ & LS & $d^6$; $t_{2g}(\uparrow)^3t_{2g}(\downarrow)^3$ & 131.4 & 3.775 & 3.775  & 0.000 & 0.500 & 0.000 & 0.500 & - \\ 
    & IS & $d^6$; $t_{2g}(\uparrow)^3t_{2g}(\downarrow)^2e_{g}(\uparrow)^1$ & 97.7 & 3.815 & 3.815  & 0.000 & 0.500 & 0.000 & 0.500 & - \\
   & HS & $d^6$; $t_{2g}(\uparrow)^3t_{2g}(\downarrow)^1e_{g}(\uparrow)^2$ & 118.1$i$ & 3.807 & 4.577 & 0.000 & 0.443 & 0.827 & 0.309 & $-79$ \\ 
  Ni$^{3+}$ & LS & $d^7$; $t_{2g}(\uparrow)^3t_{2g}(\downarrow)^3e_{g}(\uparrow)^1$ & 112.3 &  3.815 & 3.815  & 0.000 & 0.500 & 0.000 & 0.500 & -  \\ 
   & HS & $d^7$; $t_{2g}(\uparrow)^3t_{2g}(\downarrow)^2e_{g}(\uparrow)^2$ & 47.4$i$ & 3.758 & 4.632 & 0.000 & 0.453 & 0.841 & 0.318 & $-208$ \\
  Cu$^{3+}$ &  & $d^8$; $t_{2g}(\uparrow)^3t_{2g}(\downarrow)^3e_{g}(\uparrow)^2$ & 61.3$i$ &  3.830 & 4.010  & 0.000 & 0.468 & 0.938 & 0.414 & $-15$ \\  \hline\hline
\end{tabular} 
\caption{The electronic configuration (Elec. Config.) of each B-cation studied is shown, along with the frequency of the ferroelectric phonon mode ($\omega_{FE}$) for each LaBO$_3$ perovskite in the cubic phase (at $\Gamma$). An imaginary frequency suggests a ferroelectric instability, a positive frequency suggests the system is stable against a ferroelectric distortion. The relaxed lattice vectors and internal coordinates are also shown for the LaBO$_3$ systems. Internal coordinates are given in lattice units along the $c$ direction, which is the direction that ions are displaced in a ferroelectric distortion. For systems that prefer the ferroelectric phase, the energy lowering (with respect to the paraelectric phase) associated with the distortion is also presented as $\Delta E_{FE}$.}
\label{tab:SLC2}
\end{table*}

\begin{table}
\setlength{\tabcolsep}{12pt}
\setlength{\extrarowheight}{4.5pt}
\begin{tabular}{lcccc} \hline\hline
& La & Fe  & O1 & O2 \\ \hline
HS LaFeO$_3$ & $-$0.60 & 0.25 & 0.28 & 0.50 \\ 
\hline\hline
\end{tabular} 
\caption{An example of the eigenvector for the phonon mode associated with the ferroelectric lattice instability. The present calculation is for the high spin (HS) state of LaFeO$_3$.}
\label{tab:mode}
\end{table}

Looking at Table~\ref{tab:SLC2}, starting with LaScO$_3$, a strong ferroelectric instability is  present -- the ferroelectric phonon mode is softened ($\omega_{FE} = 155.7i$ cm$^{-1}$), and the system is strongly tetragonal ($c/a = 1.25$) with ferroelectric distortions of the ScO$_6$ octahedron. This result is somewhat expected as Sc$^{3+}$ has the typical $d^0$ configuration for a ferroelectric B-cation. Moving across the $3d$-series for Ti$^{3+}$, V$^{3+}$, and Cr$^{3+}$, as electrons are added to the B-cation, the ferroelectric instability begins to disappear, and as the ferroelectric mode is hardened the paraelectric structure is preferred. The observed trend with increasing occupation of the $d$ band is entirely as expected, and in accordance with the $d^0$ rule. What is somewhat surprising is that for HS $d^5-d^7$ and $d^8$ cations, the ferroelectric instability is recovered. For HS LaFeO$_3$, LaCoO$_3$, and LaNiO$_3$, and for LaCuO$_3$, the ferroelectric mode is softened, and the polar distortion becomes energetically favoured over the paraelectric phase. The HS state of LaMnO$_3$ exhibits a slight ferroelectric distortion ($c/a = 1.02$), with a total energy only slightly lower than the paraelectric phase; as described in section~\ref{sec:PJTE}, this system can be thought of as type (ii), with $K_{0} \sim K_{v}$.

\begin{figure}
\includegraphics[width=8cm]{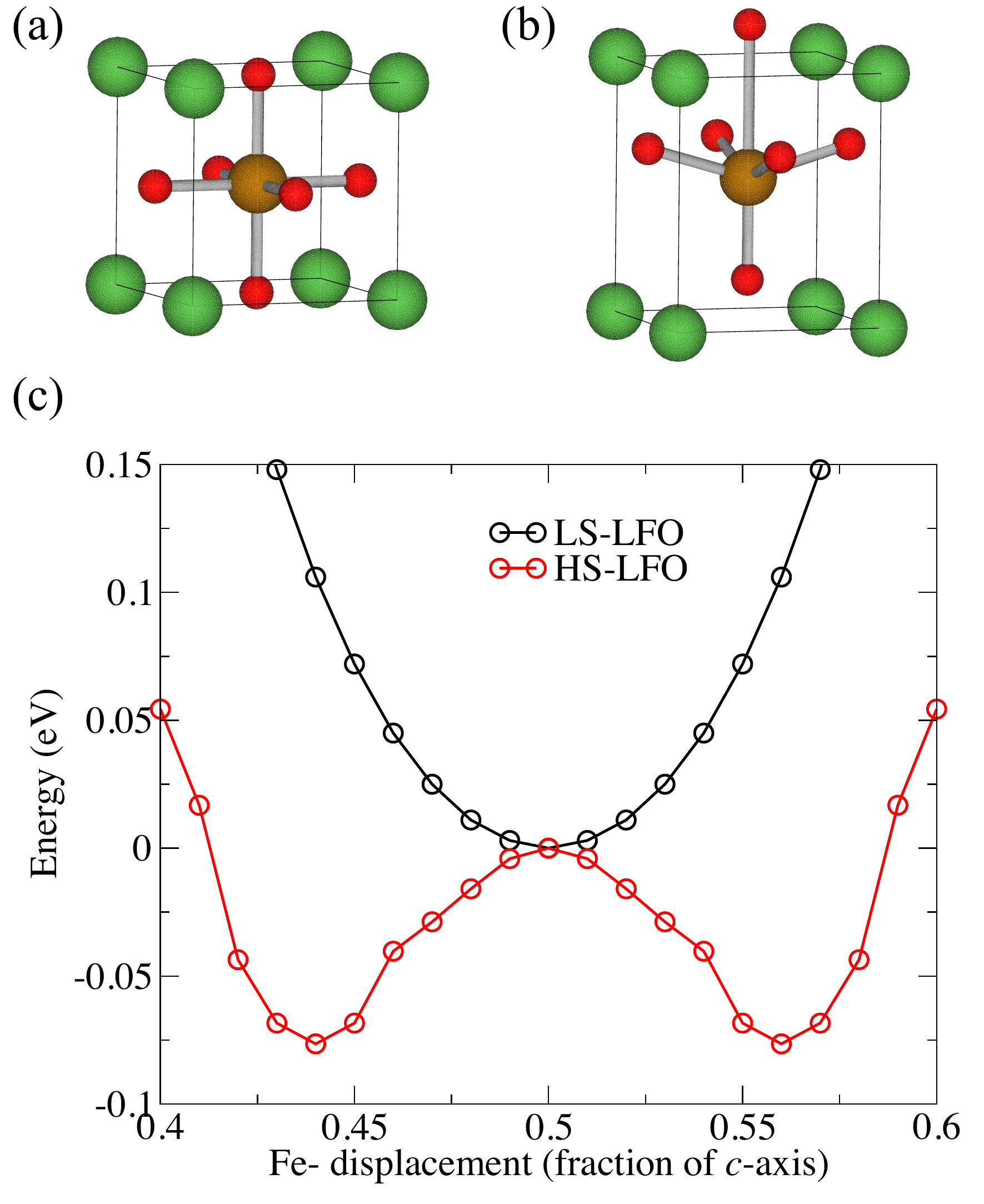}
\caption{(Color online) Crystal structure of LaFeO$_3$ (LFO) in the low spin (a) and high spin (b) states. In (c), the total energy of LaFeO$_3$ is calculated as a function of the Fe ion position in the unit cell, presented as a fraction of the $c$-axis lattice vector. Sr atoms are represented by green spheres, Fe are brown, O are red.}
\label{fig:LFO}
\end{figure}

It is noted here that the apparent coupling of the electrical polarization to the spin state is enormous -- for the $d^5-d^7$ cations, switching between spin states leads to a phase change between a paraelectric and a ferroelectric structure, i.e., these cations exhibit the multiferroic crossover effect \cite{BersukerPRL2012}. For example, in Fig.~\ref{fig:LFO} the crystal structure of LS and HS LaFeO$_3$ is shown (as obtained in Table~\ref{tab:SLC2}). In the LS state, the system is cubic, in the HS state, the unit cell is tetragonal and the atomic positions show strong polar distortions. To  further illustrate this, also in Fig.~\ref{fig:LFO}, the ferroelectric nature of LS and HS LaFeO$_3$ is demonstrated by moving the Fe ion from the central (paraelectric) position to an off-center (ferroelectric) position. The LS state of LaFeO$_3$ exhibits a single well profile with a global energy minimum at the central position. On the other hand, the HS state exhibits a deep double-well profile with minima at the off-centred positions, similar to what is displayed by other ferroelectric materials such as BaTiO$_3$ and PbTiO$_3$ \cite{CohenNAT1992}. At this point, it is somewhat counter-intuitive to think that the ferroelectric lattice instability is enhanced in the HS state of these B-cations, since here the local magnetic moments are maximal; in this case, it appears that a large number of unpaired $d$ electrons \emph{actually drives ferroelectricity}, rather than inhibits it. 

With it now established which magnetic cations exhibit a strong tendency for a ferroelectric distortion, the next step is to develop a model as to why, in terms of the microscopic and quantum mechanical effects that drive magnetism and ferroelectricity. The systematic reduction of the ferroelectric instability with orbital filling for the $d^0 - d^4$ B-cations is in accordance with the $d^0$ rule. A possible explanation for this effect has been discussed by Khomskii \cite{KhomskiiJMMM2006} in terms of two contributing factors: (i) the presence of unpaired $d$ electrons will inhibit the formation of a $p-d$ bonding spin singlet state ($\uparrow\downarrow - \downarrow\uparrow)$ due to the exchange interaction, and (ii) the population of the antibonding orbitals (with predominant $d$ character) should also weaken the $p-d$ covalent bond. However, these explanations clearly break down for the case of HS $d^5-d^7$ and $d^8$ B-cations. For example, HS Fe$^{3+}$ has 5 unpaired $d$ electrons, and therefore the influence of the``pair-breaking effect" of the exchange interaction should be very strong, and moreover there is a significant population of the antibonding states, yet LaFeO$_3$ shows a strong ferroelectric instability.

We identify that the stabilization of the polar phase for HS $d^5-d^7$ and $d^8$ B-cations can be understood within the PJT theory for ferroelectricity; here, note that we refer to ``stabilization" with respect to the cubic phase, not the ground state structures presented in Table~\ref{tab:structure}. According to Eq.~\ref{PJTEsum}, the lattice is stabilized against a ferroelectric distortion via the $K_0$ term. However, this term is expected to be strongly affected by an increase in the volume of the unit cell \cite{GarciaIC2014}. Indeed, looking at the lattice constants presented in Table~\ref{tab:SLC2}, for the HS states of the $d^5-d^7$ and $d^8$ LaBO$_3$ perovskites the volume is significantly increased in the HS state. For example, LaFeO$_3$ undergoes a 23 \% increase in the volume in the HS state, when compared to the LS state. This dramatic increase in the volume is due to the population of the antibonding ($\sigma^*$) $e_g$ states. Recently, Garcia et al.\ \cite{GarciaIC2014} have shown that the local forces that stabilize against a PJT driven ferroelectric distortion are strongly reduced by increasing the volume. As an example, to determine quantitatively the role of the volume increase in stabilizing the polar phase of the HS state in the LaBO$_3$ systems, the frequency of the ferroelectric phonon mode $\omega_{FE}$ for the LS state of LaFeO$_3$ has been calculated, however the volume has been fixed to the equilibrium volume of HS LaFeO$_3$. Indeed, at the HS volume, the mode is softened significantly with $\omega_{FE} = 72.1i$. This does indeed suggest a significant decrease in $K_0$ for the increased volume. Keeping the same volume, and switching to the HS state, the mode is softened slightly further, with $\omega_{FE} = 87.8i$. The discrepancy in $\omega_{FE}$ could be attributed to an increase in the contribution from $K_v$. To further investigate the effect of the volume increase, we also performed relaxations for the LS state fixed at the volume of the HS state; indeed, now a polar phase is preferred with respect to the cubic structure. The $c/a$ ratio is 1.17 and the system exhibits internal ferroelectric distortions, further supporting the idea that a reduction in the $K_0$ term stabilizes the polar state. The $c/a$ ratio and the internal distortions are slightly less than those of the HS state for the same volume, which could also point to a contribution from $K_v$; however, it seems that the effect of volume on $K_0$ is particularly strong.

\begin{figure}
\includegraphics[width=8cm]{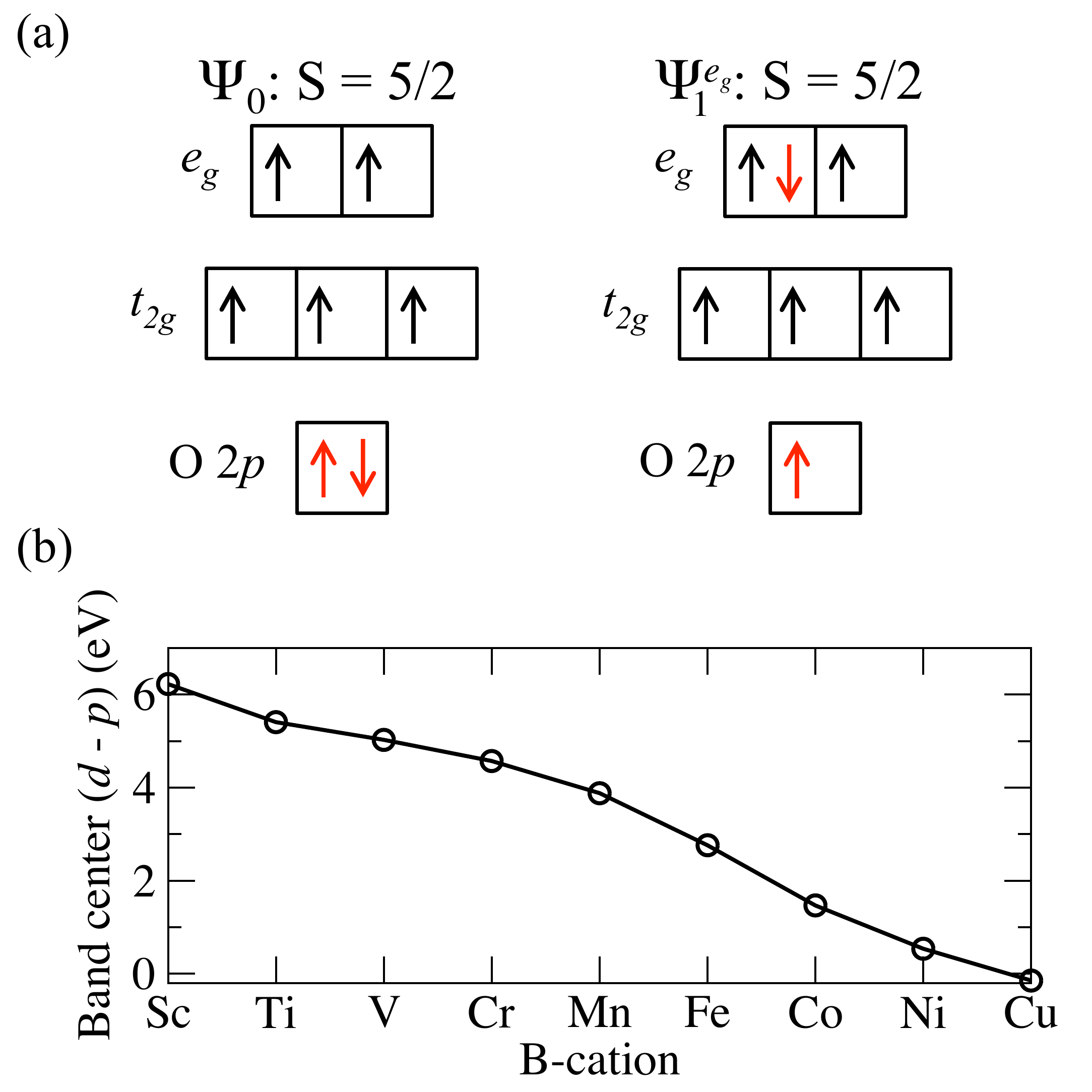}
\caption{(Color online) (a) On the left the ground state orbital configuration ($\Psi_0$) is shown for high spin Fe$^{3+}$ in LaFeO$_3$. On the right, the same orbital configuration is shown however an electron has been promoted from the O $2p$ band into the $e_g$ band of Fe$^{3+}$, the label $\Psi_1^{e_g}$ indicates that this is the lowest energy excited state involving promotion of an electron into the $e_g$ band. As can be seen $\Psi_0$ and $\Psi_1^{e_g}$ have the same spin $S = 5/2$, and so $\Psi_1^{e_g}$ contributes to the summation in Eq.~\ref{PJTEvibronic}. (b) The energy difference between the O $2p$ and B-cation $3d$ band centres is plotted for the LaBO$_3$ series. The position of the band center is approximated \cite{note1}, and this plot should only be considered as a qualitative guide.}
\label{fig:PJTE}
\end{figure}

With regard to the $K_v$ term, as discussed in section~\ref{sec:PJTE}, from Eq.~\ref{PJTEvibronic}, it is clear that the strength of the vibronic coupling between a $p$ and $d$ orbital depends on several factors: (i) whether or not the promotion of an electron from the occupied $p$ state into the unoccupied (or partially occupied) $d$ state conserves the spin of the system \cite{BersukerPRL2012}, (ii) the matrix element, $\braket{p |  \delta H/\delta q | d}$, which describes whether or not the distortion increases the $p-d$ overlap, and (iii) the energy gap between the two orbitals ($E_p - E_d$). It is this dependence that leads to the so-called multiferroic crossover effect, whereby different orbital configurations lead to an enhanced contribution from the vibronic term \cite{BersukerPRL2012}. Looking at HS LaFeO$_3$, the ground state $\Psi_0$ has an orbital configuration with $S = 5/2$; an example excitation $\Psi_n$ involving the $e_g$ band, that conserves the spin of the system, is shown in Fig.~\ref{fig:PJTE}. In this case, the excited state has the same spin as $\Psi_0$, and so contributes to the summation in Eq.~\ref{PJTEvibronic}. It follows, that the number of $\Psi_n$ terms that contribute to the summation in Eq.~\ref{PJTEvibronic} (and the strength of the contribution) is dependent on the orbital configuration, i.e., the multiferroic crossover effect \cite{BersukerPRL2012}, and therefore the $p - d$ vibronic interaction ($K_v$) is enhanced in one spin state with respect to another.

\begin{figure}
\includegraphics[width=8cm]{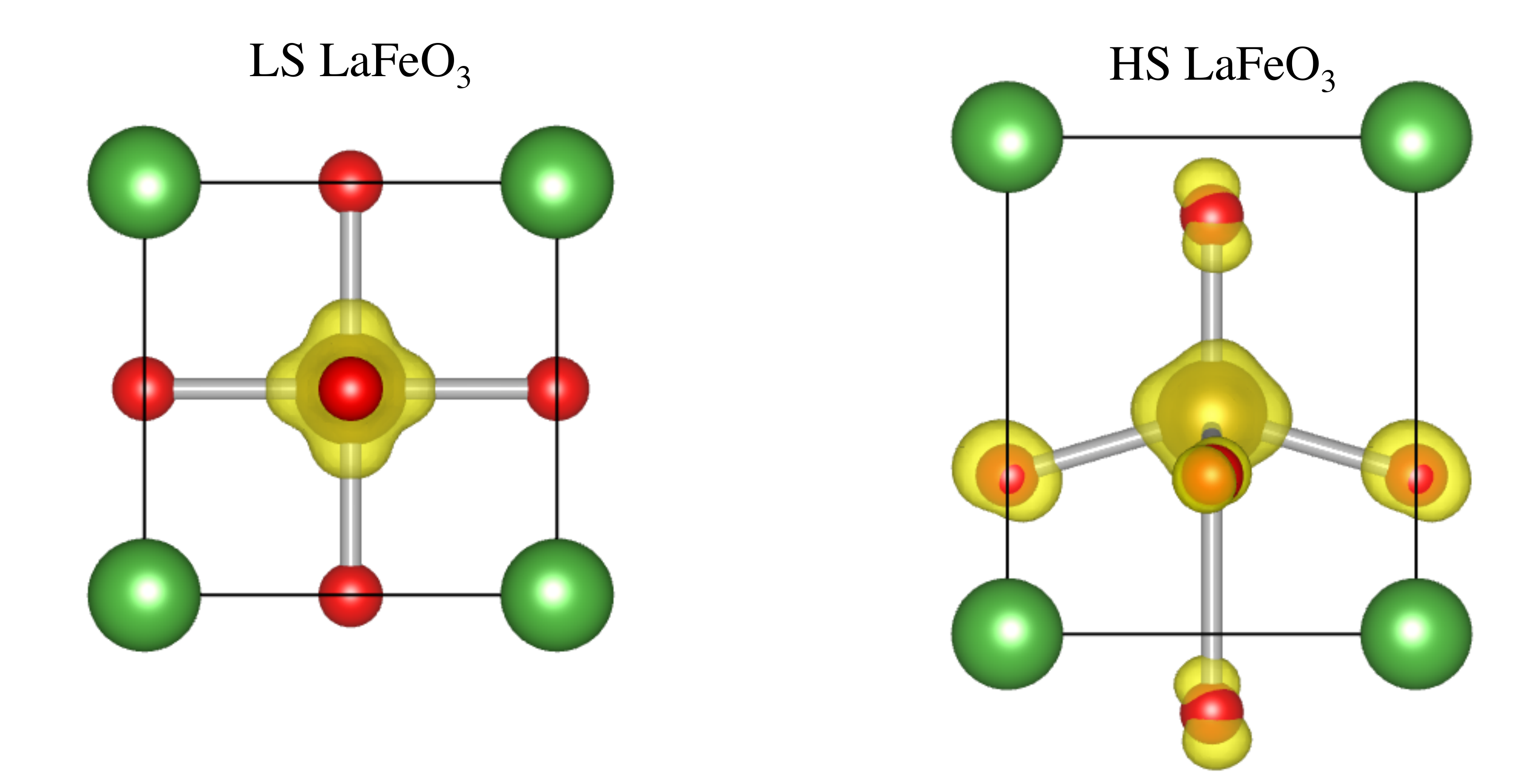}
\caption{(Color online) The charge density isosurface is shown for the $e_g$ band of low spin (LS) (left) and high spin (HS) (right) LaFeO$_3$. The isosurface is set to 10\% of the maximum value. La ions are prepresented by green spheres, Fe are brown and O are red.}
\label{fig:iso}
\end{figure}

To confirm that the $\sigma$-bonding $p-e_g$ interaction is enhanced in the HS state, in Fig.~\ref{fig:iso} we plot the charge density isosurface for the $e_g$ states of the Fe$^{3+}$ ion in the LS and HS states of LaFeO$_3$. In the LS state, the charge density for the $e_g$ molecular orbitals are localized on the Fe$^{3+}$ ion, without significant hybridization with the surrounding O $2p$ states. On the other hand, in the HS state, the $e_g$ band is hybridized and has significant O $2p$ character; this is a clear indication of an increase in the PJT $p-d$ covalent interaction. A similar analysis shows an increased in the $p - t_{2g}$ hybridization. We are able to approximate the total increase in covalent bonding based on the $\Gamma$-point projections of the bonding valence states onto the Fe and O atoms. In the LS state, the Fe $3d$ states contribute approximately 8\% to the bonding states of the predominantly O $2p$ valence band. When switching to the HS state, indeed the bonding states are now comprised of 16\% Fe 3$d$ character, confirming increased $p-d$ hybridization. Therefore, to summarise, we propose that the recovery of a strong ferroelectric instability for HS $d^5-d^7$ and $d^8$ B-cations is due to a significant reduction of the $K_0$ term due to the volume increase, as well as increased $p-d$ bonding ($K_v$). 

As an additional point, we notice that the HS $d^5-d^7$ and $d^8$ B-cations cations all lie to the right of the $3d$-block. LaScO$_3$ has a $p-3d$ band gap of nearly 5 eV (Fig.~\ref{fig:pdos}), however moving to the right of the $d$-block, the increased nuclear charge should bind the $d$-electrons more tightly, and reduce the $p-d$ energy gap; this would reduce the denominator in Eq.~\ref{PJTEvibronic}, and increase the $p-d$ vibronic interaction. In Fig.~\ref{fig:PJTE}, we plot the energy difference between the band center of the O $2p$ band and that of the B-cation $3d$ band for the LaBO$_3$ systems \cite{note1}. As can be seen, to the left of the $3d$-block the average energy of the B-cation $3d$ band lies at a much higher energy than the O $2p$ band. However, moving to the right of the $3d$-block the energy gap is systematically reduced, and for $d^5-d^7$ and $d^8$ cations the O $2p$ and B-cation $3d$ bands lie at a similar energy such that the potential for PJT driven ferroelectricity is greatly increased. However, we point out that, clearly, a small $p-d$ energy gap alone is insufficient to drive a ferroelectric distortion, as the condition $K_v > K_0$ must be satisfied.

The above discussion for the LaBO$_3$ perovskites should be quite generalizable for these B-cations if the A-cation is replaced. To confirm this, we studied the ferroelectric properties of LS and HS CeFeO$_3$. This allows us to confirm the key findings from above, namely: (i) a strong ferroelectric instability for a half-filled $e_g$ band (which is absent for other orbital configurations), and (ii) the strong coupling of the ferroelectric polarization to the spin state. Indeed, we find that for HS CeFeO$_3$ the ferroelectric phonon mode is softened with $\omega_{FE} = 97.5i$ cm$^{-1}$, and the relaxed structure is tetragonal with $c/a = 1.24$ and strong polar distortions; for LS CeFeO$_3$ the ferroelectric mode is hardened to $\omega_{FE} = 112.5$ cm$^{-1}$ and the system relaxes back the paraelectric phase. Therefore, the behaviour of Fe$^{3+}$ in LaFeO$_3$ is well reproduced in CeFeO$_3$. We emphasise that, again, CeFeO$_3$ does not crystallize in the tetragonal ferroelectric structure; rather, this system prefers an orthorhombic distortion \cite{YuanJAP2013}. However, the previous result illustrates that the ferroelectric and multiferroic properties presented for LaFeO$_3$ arise due to the $d^5$ Fe$^{3+}$B-cation, and not some other effect as a result of the A-cation. 

At this point it has been demonstrated that LaFeO$_3$, LaCoO$_3$ and LaNiO$_3$ can potentially exhibit the multiferroic crossover effect, and this strong coupling of the electrical and magnetic polarization is highly desirable from a multiferroic device perspective. However, unfortunately, these materials prefer to undergo centrosymmetric distortions, rather than a polar one. Specifically, at room temperature LaCoO$_3$ and LaNiO$_3$ exhibit rhombohedral distortions \cite{BiswasJAC2009, ThorntonJSSC1986}, whereas LaFeO$_3$ is orthorhombic \cite{MarezioMRB1971}. The importance of competitive distortions in the $d^0$ rule has already been stressed previously \cite{HillJPC2000}. Such distortions can originate from the JT effect, or, in the absence of a JT distortion, the PJT effect can also give rise to centrosymmetric distortions \cite{GarciaJCPL2010}. One possible route to suppressing these competing distortions is via strain. Tensile strain is known to inhibit orthorhombic and rhombohedral distortions, and favours a polar tetragonal phase  \cite{SchlomARMR2007}. It is therefore possible that through substrate engineering of thin films, the LaBO$_3$ systems studied above could be stabilized in the polar tetragonal phase via strain. Indeed, it was shown that strain engineering could be used to stabilize a weak ferroelectric state in magnetic AMnO$_3$ perovskites \cite{RondinelliPRB2009, BhattacharjeePRL2009}. A second approach to suppressing the competing centrosymmetric distortions is via the choice of the A-cation. It is a well known empirical fact that the tilting and rotation of BO$_6$ octahedra depends strongly on the relative sizes of the A and B cations  \cite{GoldsmithNAT1926}. A larger A-cation compared to a smaller B-cation will inhibit these distortions. For perovskites of the A$^{3+}$B$^{3+}$ charge ordering, La$^{3+}$ is already very large, and replacing the A-cation with most of the alternative 3$+$ cations (e.g., other the rare earth ions) would increase the tendency for competing centrosymmetric distortions. One possibility is to focus on A$^{2+}$B$^{4+}$O$_3$ systems, as $2+$ cations are expected to be larger -- indeed, the well known ferroelectric perovskites PbTiO$_3$ and BaTiO$_3$ have the A$^{2+}$B$^{4+}$ charge ordering. Surprisingly, there has been far less work on magnetism, or multiferroicity, in the A$^{2+}$B$^{4+}$O$_3$ perovskites. Focusing on $d$-fillings which can lead to the multiferroic crossover effect, i.e., $d^5 - d^7$, and taking into account that Cu will not form in the 4$+$ charge state, the possible B-site cations are Co$^{4+}$ and Ni$^{4+}$. For large A-cations in the 2$+$ charge state Sr$^{2+}$, Ba$^{2+}$ and Pb$^{2+}$ are plausible candidates. A survey of the literature suggests that in most cases the different combinations of these A- and B-cations will result in a ground state structure in the hexagonal phase  \cite{TakedaJINC1972, GottschallIC1998, InoueIEEE2005}, which commonly occurs when an A-cation is much larger than the B-cation. For the case of SrCoO$_3$, it was indeed reported that the centrosymmetric distortions were suppressed, and the cubic structure is the ground state phase   \cite{LongJPCM2011}. However, SrCoO$_3$ has an IS ground state with a $t_{2g}^4e_g^1$ orbital filling which, as we have shown, does not lead to a strong ferroelectric instability.

\subsection{Multiferroic Crossover: BiCoO$_3$}

\label{sec:MFCE}

\begin{figure}
\includegraphics[trim=0cm 1.5cm 3cm 1.5cm,clip=true,width=8.5cm]{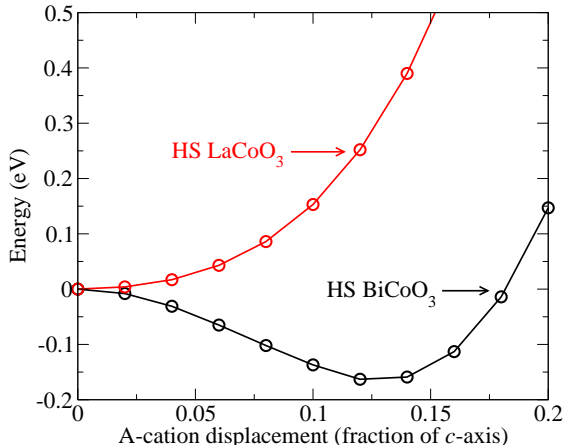}
\caption{(Color online) The total energy (per formula unit) of cubic high spin (HS) BiCoO$_3$ and cubic HS LaCoO$_3$ is plotted as a function of the A-cation displacement from the high symmetry site.}
\label{fig:Bi-La}
\end{figure}

In the previous section of the manuscript, it was revealed that certain $d^n$ cations can indeed provide the driving force for a ferroelectric distortion; however, competition with noncentrosymmetric distortions can account for the empirical $d^0$ rule \cite{HillJPC2000, BenedekJPC2013}. The perovskite BiCoO$_3$ appears to be an exception to this rule, as the ground state crystal structure is that of a tetragonal ferroelectric phase, and yet the $d^6$ Co$^{3+}$ ion is in the HS state with $C$-type antiferromagnetic ordering \cite{BelikCM2006} (for more detailed information on this system see Ref.~[\onlinecite{JiaPRB2012}]). It was suggested by Jia et al.\ that hybridization of the high lying Bi $6s$ state with the O $2p$ valence band stabilizes the ferroelectric distortion \cite{JiaPRB2012}. Indeed, surely the Bi ion will play a role. To demonstrate this, in Fig.~\ref{fig:Bi-La}, we plot the total energy of cubic HS BiCoO$_3$ as a function of the displacement of the Bi ion from the high symmetry position. For comparison, we also plot the same result for cubic HS LaCoO$_3$. Clearly, while the total energy of LaCoO$_3$ is increased by the displacement, the total energy of BiCoO$_3$ is reduced, consistent with Bi $6s$ -- O $2p$ hybridization \cite{JiaPRB2012}. However, based on the results of Table~\ref{tab:SLC2}, it seems likely that the HS state of a $d^6$ ion provides an additional driving force for the ferroelectric distortion via PJT vibronic coupling. Indeed, in a previous report, we have demonstrated the multiferroic crossover effect for Co$^{3+}$ dopants at the Ti site in PbTiO$_3$ \cite{WestonPRL2015}. In the report by Jia et al. \cite{JiaPRB2012}, the authors fixed the volume of their BiCoO$_3$ cell to that of the experimental structure, and restricted the Co$^{3+}$ ions to be in the LS state; in this case the system favours a tetragonal phase, and based on this result, the authors concluded that the distortion is due purely to Bi -- O hybridization, ruling out a contribution from Co$^{3+}$. However, clearly, by fixing the volume of the system to that of the experimental structure, the effect of the reduced $K_0$ term caused by the LS -- HS spin transition cannot be observed; additionally, the increase in the vibronic term for HS Co$^{3+}$ was not explored.

\begin{figure}
\includegraphics[trim=3.5cm 1.5cm 4cm 0.8cm,clip=true,width=8cm]{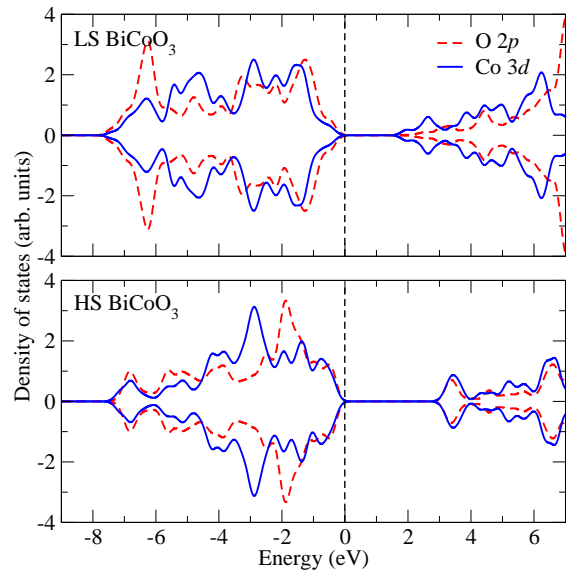}
\caption{(Color online) The partial density of states (DOS) for low spin (LS) (top) and high spin (HS) (bottom) BiCoO$_3$. The LS state of BiCoO$_3$ is non-magnetic, whereas the HS state has $C$-type antiferromagnetic ordering. The DOS has been shifted so that the Fermi level is at zero.}
\label{fig:BCO}
\end{figure}

\begin{figure}
\includegraphics[width=8.5cm]{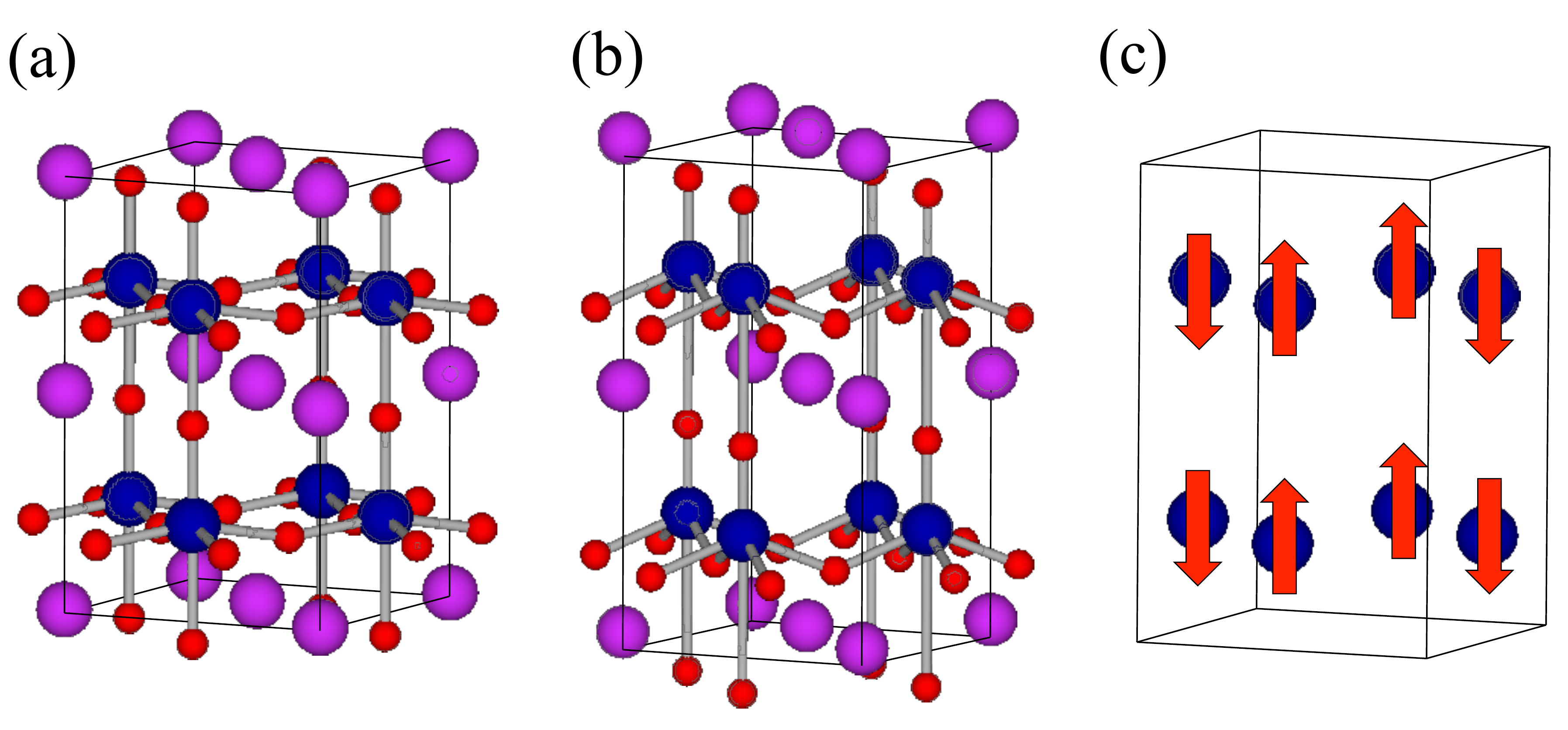}
\caption{(Color online) The calculated crystal structure for BiCoO$_3$ in the low spin (a) and high spin (b) states. In (c), the ground state $C$-type antiferromagnetic ordering of the high spin Co ions is shown. Bi atoms are represented by purple spheres, Co are blue and O are red.}
\label{fig:BCO-structure}
\end{figure}

\begin{table}
\setlength{\tabcolsep}{5pt}
\setlength{\extrarowheight}{4.5pt}
\begin{tabular}{lcccccc} \hline\hline
 Spin state & $a$ & $c$ & Bi & Co & O1 & O2 \\ \hline
 LS  & 3.731 & 4.056 & 0.000 & 0.426 & 0.881  & 0.360 \\ 
 HS  & 3.727 & 4.825 & 0.000 & 0.424 & 0.892  & 0.269  \\   \hline\hline
\end{tabular} 
\caption{Lattice vectors and internal coordinates of BiCoO$_3$ in the low spin (LS) and high spin (HS) states.}
\label{tab:BCO}
\end{table}

To determine the role of the HS Co$^{3+}$ ion in driving the ferroelectric state of BiCoO$_3$, we have investigated the atomic and electronic structure of BiCoO$_3$ using accurate hybrid functional calculations. Our results confirm that BiCoO$_3$ has a HS configuration as the ground state with $C$-type antiferromagnetic ordering, and a strongly tetragonal ($c/a = 1.29)$ crystal structure with large internal ferroelectric distortions. Our calculated crystal structure for the LS and HS states of BiCoO$_3$ is shown in Fig.~\ref{fig:BCO-structure}; the $C$-type antiferromagnetic ordering is also shown. A $\sqrt{2}a \times \sqrt{2}b \times 2c$, 20-atom BiCoO$_3$ cell is used to account for the long range magnetic ordering. The lattice vectors and internal coordinates calculated for the LS and HS sates of BiCoO$_3$ are presented in Fig.~\ref{tab:BCO}. Importantly, we find that in the LS state the axis ratio is reduced to $c/a = 1.08$, and the internal ferroelectric distortions are reduced; this result confirms the multiferroic crossover effect for BiCoO$_3$, and moreover, this result reveals the crucial role of the Co$^{3+}$ ion in driving the ferroelectric state, which has been previously attributed to Bi $6s$ -- O $2p$ hybridization. It is noted, however, that unlike the LaBO$_3$ systems, BiCoO$_3$ exhibits a remnant polarization in the LS state, supporting the idea that the Bi $6s$ does indeed provide a stabilizing effect for the ferroelectric distortion.

As the ferroelectric polarization of BiCoO$_3$ is affected by switching the spin state, potentially, manipulation of the polarization -- e.g., by strain, or by an externally applied electric field -- could induce spin crossover. This idea is supported by recent first-principles calculations based on the generalized gradient approximation (GGA)  \cite{RavindranAM2008}, where the paraelectric phase of BiCoO$_3$ was reported to prefer the LS state. However, the LS state was found to be metallic which would prevent switching via an applied electric field. It should be pointed out though, that these predictions for the electronic and magnetic structure should not be considered reliable as the GGA cannot accurately predict the band gaps of semiconductors, and because the GGA will overly favour the LS state due to the large self-interaction error \cite{ReiherTCA2001}.

\begin{figure}[t]
\includegraphics[trim=0.5cm 1.5cm 6cm 2.5cm,clip=true,width=8cm]{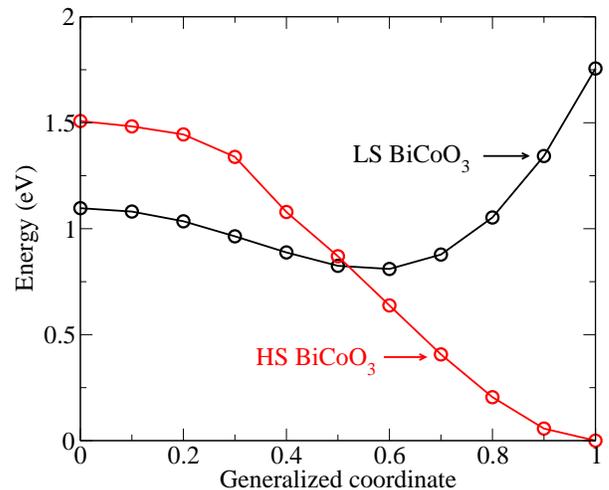}
\caption{(Color online) The total energy (per formula unit) of low spin (LS) and high spin (HS) BiCoO$_3$ is plotted as a function of the generalized coordinate. To model the dependence of the spin state energetics on the polarization, the internal coordinates are moved continuously from those of the relaxed high spin structure (generalized coordinate = 1) to a paraelectric structure (generalized coordinate = 0).}
\label{fig:SWITCH}
\end{figure}

Using advanced hybrid functional calculations, in Fig.~\ref{fig:BCO} we are able to confirm that BiCoO$_3$ is semiconducting in \emph{both} spin states, with band gaps of 2.3 eV and 1.2 eV for HS and LS BiCoO$_3$, respectively. This result suggests that an applied electric field could be used to manipulate the internal polarization. The hybrid functional calculations are also able to give an accurate treatment of the HS -- LS splitting of spin crossover systems \cite{ReiherTCA2001, HarveyBOOK2004}, and therefore we are able to more accurately predict the way in which changes in the electric polarization will affect the spin state energetics. An applied field can reduce the internal polarization of a ferroelectric material \cite{SaiPRB2002}, and force the internal coordinates to be closer to that of the paraelectric phase. In Fig.~\ref{fig:SWITCH}, the dependence of the spin state energetics on the ferroelectric polarization is presented for BiCoO$_3$. The internal coordinates are fixed by interpolating between that of the relaxed HS geometry and the paraelectric phase, however at each point we allow relaxation of the lattice vectors. As can be seen, at the polar HS geometry, the total energy of the HS state is far lower than that of the LS state. As the internal polarization is reduced, the HS -- LS splitting is reduced and at about the halfway point of the interpolation, the LS state becomes the ground state (note that we also performed test calculations for other magnetic configurations of the HS state, and also for the IS state to confirm that the LS state becomes the ground state). Figure~\ref{fig:SWITCH} provides a good first indication for electric field control of magnetism in BiCoO$_3$, i.e., changes in the internal polarization strongly affect the relative energies of the HS and LS states; we propose experimental studies to confirm this magnetoelectric effect.

\section{Summary}
\label{sec:summary}
In this report, using the LaBO$_3$ series as a model, we have performed a qualitative trend study so as to investigate the interaction between ferroelectricity and magnetism in ABO$_3$ perovskites. For the B-cations studied, moving to the right of the $3d$-series, it was found that initially, increasing the occupation of the B-cation $d$ orbital decreases the tendency for a ferroelectric distortion, which is in agreement with the ferroelectric $d^0$ rule. However, a surprising result was found for HS $d^5-d^7$ and $d^8$ cations, in that a strong ferroelectric instability was recovered. This result was explained in terms of the pseudo Jahn-Teller theory for ferroelectricity, and it was demonstrated that, contrary to the current understanding of the ferroelectric $d^0$ rule, in some cases unpaired $d$ electrons actually \emph{drive} ferroelectricity, rather than inhibit it. For the case of BiCoO$_3$, the crucial role of the Co$^{3+}$ ion in driving the ferroelectric lattice instability has been revealed. Finally, it was demonstrated that $d^5-d^7$ B-cations will exhibit the multiferroic crossover effect, whereby switching between spin states can strongly affect the ferroelectric polarization; moreover, the manipulation of the polarization will also affect the relative energies of the different spin states, which suggests the possibility of electric field control of magnetism.


\begin{thebibliography}{58}
\bibitem[{not({\natexlab{a}})}]{noteCHARGE}
\bibinfo{note}{Formal charge only; this convention will be used throughout the
  manuscript.}
\bibitem[{not({\natexlab{b}})}]{note1}
\bibinfo{note}{The $d$-band center is determined using non-spin-polarized
  calculations of the cubic phase for each LaBO$_3$ perovskite. The
  integrations are done from the bottom of the O $2p$ band to the top of the
  B-cation $3d$ band. This approach leads to some error in the calculation due
  to covalent bonding, however the approach is adequate to show the qualitative
  trends in the energy gap between the O $2p$ and B-cation $3d$ bands.}
\end{thebibliography}

\end{document}